\documentstyle[aps,multicol,prb]{revtex}

\begin{document}

\title{Comment on ``Fun and frustration with
quarkonium in a $1+1$ dimension,'' by R.\ S.\ Bhalerao
and B.\ Ram [Am.\ J.\ Phys.\ {\bf 69} (7), 817--818 (2001)]}
\author{R.\ M.\ Cavalcanti\cite{email}}
\address{Instituto de F\'{\i}sica, Universidade Federal
do Rio de Janeiro \\
Caixa Postal 68528, 21945-970 Rio de Janeiro, RJ, Brazil}
\maketitle

\begin{multicols}{2}

In a recent paper, Bhalerao and Ram\cite{B&R} solved
the Dirac equation in $1+1$ dimensions with a Lorentz
scalar potential given by
\begin{equation}
V(x)=g|x| \qquad(g>0)
\end{equation}
and came to the conclusion that the energy levels
are given by
\begin{equation}
\label{spec}
E=\pm\sqrt{2(n+1)g}  \qquad(n\in{\cal I}),
\end{equation}
where ${\cal I}$ is the set of non-negative integers
that satisfy one of the conditions
\begin{equation}
\label{cond1}
H_{n+1}(m/\sqrt{g})
=\pm\sqrt{2(n+1)}\,H_n(m/\sqrt{g}).
\end{equation}
$H_n(z)$ is the Hermite polynomial of 
degree $n$ and $m$ is the fermion mass.
The purpose of this Comment is to point out that
the energy spectrum obtained by in Ref.~\onlinecite{B&R}
is not correct, and to derive the correct one.

Our starting point is Eq.~(6a) of Ref.~\onlinecite{B&R}.
Its general solution is given by\cite{Lebedev}
\begin{equation}
\label{psi1>}
\psi_1(\xi)=C H_{\nu}(\xi)\,e^{-\xi^2/2}
+D H_{-\nu-1}(i\xi)\,e^{\xi^2/2},
\end{equation}
where $\nu\equiv E^2/2g$, $\xi\equiv(m+gx)/\sqrt{g}$, and
$H_{\nu}(z)$ is the Hermite function,\cite{remark} defined as
\begin{eqnarray}
\label{Hermite}
H_{\nu}(z)&=&\frac{2^{\nu}\,\Gamma\left(\frac{1}{2}\right)}
{\Gamma\left(\frac{1-\nu}{2}\right)}\,\Phi\left(-\frac{\nu}{2},
\frac{1}{2};z^2\right)
\nonumber \\
& &+\frac{2^{\nu}\,
\Gamma\left(-\frac{1}{2}\right)}{\Gamma\left(-\frac{\nu}{2}\right)}\,
z\Phi\left(\frac{1-\nu}{2},\frac{3}{2};z^2\right) .
\end{eqnarray}
The confluent hypergeometric function $\Phi(a,b;z)$ is expressed as
\begin{equation}
\Phi(a,b;z)=\sum_{k=0}^{\infty}\frac{(a)_k}{(b)_k}\,
\frac{z^k}{k!}\,,
\end{equation}
with $(a)_0=1$, 
$(a)_k=\Gamma(a+k)/\Gamma(a)=a(a+1)\cdots(a+k-1)$,
$k=1,2,\ldots$

For large $z$ and fixed $\nu$, $H_{\nu}(z)$ 
behaves asymptotically as
\begin{equation}
H_{\nu}(z)\sim(2z)^{\nu}\qquad (|z|\to\infty),
\end{equation}
provided that $|{\rm arg}\,z|\le\frac{3\pi}{4}-\delta$.
Therefore, in order that $\psi_1(x)$ be square integrable
at $x\to\infty$, it is necessary and sufficient to
take $D=0$ in Eq.~(\ref{psi1>}),
{\em regardless of the value of} $\nu$. 
{}From Eq.~(5a) of Ref.~\onlinecite{B&R}
and the identity $H_{\nu}'(\xi)=2\nu H_{\nu-1}(\xi)$,
it then follows that the second component of the Dirac
bispinor $\psi$ is given by
\begin{equation}
\label{psi2>}
\psi_2(\xi)=C\,\frac{E}{\sqrt{g}}\, 
H_{\nu-1}(\xi)\,e^{-\xi^2/2}.
\end{equation}

Equations~(\ref{psi1>}) (with $D=0$) and (\ref{psi2>}) are
valid for $x\ge 0$; for $x\le 0$ they must be replaced
by
\begin{mathletters}
\begin{eqnarray}
\psi_1(\xi')&=&C'\frac{E}{\sqrt{g}}\,
H_{\nu-1}(\xi')\,e^{-{\xi'}^2/2},
\\
\psi_2(\xi')&=&C'H_{\nu}(\xi')\,e^{-{\xi'}^2/2},
\end{eqnarray}
\end{mathletters}
where $\xi'\equiv(m-gx)/\sqrt{g}$. The continuity of $\psi_1$ and
$\psi_2$ at $x=0$ leads to the condition
\begin{equation}
\label{cond2}
H_{\nu}^2\left(m/\sqrt{g}\right)
-2\nu H_{\nu-1}^2\left(m/\sqrt{g}\right)=0.
\end{equation}
This equation is equivalent to Eq.~(15) of Ref.~\onlinecite{B&R}
[Eq.~(\ref{cond1}) of this Comment]
if $n$ is not required to be a non-negative integer.
In Ref.~\onlinecite{B&R}, however, this equation
is a supplementary condition imposed on the previously
determined energy levels (\ref{spec}). Here, in contrast,
the energy levels are determined solely by Eq.~(\ref{cond2}). 

It remains to be seen whether Eq.~(\ref{cond2}) has any
solutions. In the Appendix, I prove that
it has an infinite number of solutions if $m=0$.
Unfortunately, I have not been able to generalize the
proof to the $m \neq 0$ case. However, a numerical attack to the
problem reveals the existence of solutions
to Eq.~(\ref{cond2}) for $m\ne 0$. Table \ref{T1}
lists the first five solutions to that equation 
for several values of $m$. 

\begin{table}

\caption{First five values of $\nu\equiv E^2/2g$ 
for different values of $\alpha\equiv m/\sqrt{g}$.}
\label{T1}

\begin{tabular}{c c c c}
$\nu$ & $\alpha=0$ & $\alpha=1$ & $\alpha=2$ \\
\hline
$\nu_0$ & 0.345459 & 1.396274 & 3.338595 \\
$\nu_1$ & 1.548571 & 3.056760 & 5.452161 \\
$\nu_2$ & 2.468573 & 4.306277 & 7.006087 \\
$\nu_3$ & 3.522295 & 5.615211 & 8.568946 \\
$\nu_4$ & 4.482395 & 6.804771 & 9.978608 \\
\end{tabular}

\end{table}

\acknowledgments

I thank Dr.\ R.\ S.\ Bhalerao for his 
comments on the first version of this paper. 
This work was supported by FAPERJ.

\appendix
\section{}
Let me show that Eq.~(\ref{cond2})
has an infinite number of solutions if $m=0$.
Indeed, in this case Eq.~(\ref{cond2}) reduces to
\begin{equation}
f(\nu)\equiv H_{\nu}^2(0)-2\nu H_{\nu-1}^2(0)=0.
\end{equation}
It is clear that $\nu$ cannot be negative,
for then $f(\nu)$ is strictly positive. If $\nu=0$ or $\nu=2k$,
$k$ a positive integer, then
$2\nu H_{\nu-1}^2(0)$ vanishes (in the latter
case because $H_{2k-1}(x)$ is an odd function of $x$) and
$H_{\nu}^2(0)>0$. Therefore, $f(\nu)>0$
if $\nu$ is a non-negative even integer.
On the other hand, $f(\nu)<0$ if
$\nu$ is a positive odd integer,
for then $H_{\nu}(0)=0$ and
$2\nu H_{\nu-1}^2(0)>0$. Because 
$f(\nu)$ is a continuous function of $\nu$,
one is led to the conclusion that $f(\nu)$ vanishes
at least once in each interval $(n,n+1)$, $n=0,1,2,\ldots$
A more refined analysis, in which one uses the identity 
(cf.\ Eq.~(\ref{Hermite}))
\begin{equation}
H_{\nu}(0)=\frac{2^{\nu}\,\Gamma\left(\frac{1}{2}\right)}
{\Gamma\left(\frac{1-\nu}{2}\right)}\,,
\end{equation}
shows that $f(\nu)$ in fact
vanishes only once in these intervals.
It follows that the energy levels
in the massless case are given by
\begin{equation}
E=\pm\sqrt{2\nu_ng}\,,
\end{equation}
with $\nu_n$ satisfying $n<\nu_n<n+1$, $n=0,1,2,\ldots$

\end{multicols}


\begin{references}
\bibitem{remark} The Hermite function $H_{\nu}(z)$ should not
be confused with the Hermite polynomials. The latter are a
special case of the former, when $\nu$ is a non-negative integer.
\end{references}
\end{document}